\documentclass[conference]{IEEEtran}
\IEEEoverridecommandlockouts
\usepackage{cite}
\usepackage{hyperref}
\usepackage{amsmath,amssymb,amsfonts}
\usepackage{algorithmic}
\usepackage{graphicx}
\usepackage{textcomp}
\usepackage{adjustbox}
\usepackage[table,xcdraw]{xcolor}
\usepackage{xcolor}
\usepackage{paralist}
\usepackage{url}
\usepackage{amsmath}
\usepackage{mathtools}
\usepackage{multicol}
\usepackage{amssymb}
\usepackage{subcaption,lipsum,booktabs}
\usepackage{tkz-graph}
\usepackage{tikz}
\usepackage{supertabular}
\usepackage{caption, subcaption}

\usepackage{amsfonts}
\usepackage{graphicx}
\usepackage{paralist}
\usepackage{rotating}
\usepackage{booktabs}
\usepackage{diagbox}
\usepackage{multirow}
\usepackage{algorithm}
\usepackage{enumitem}
\usepackage{algorithmic}
\setlist[itemize]{noitemsep, topsep=0pt}
\setlist[enumerate]{noitemsep, topsep=0pt}
\def\BibTeX{{\rm B\kern-.05em{\sc i\kern-.025em b}\kern-.08em
    T\kern-.1667em\lower.7ex\hbox{E}\kern-.125emX}}

\usepackage{listings}

\colorlet{punct}{red!60!black}
\definecolor{background}{HTML}{EEEEEE}
\definecolor{delim}{RGB}{20,105,176}
\colorlet{numb}{magenta!60!black}

\lstdefinestyle{base}{
  language=C,
  emptylines=1,
  breaklines=true,
  basicstyle=\ttfamily\color{black},
  moredelim=**[is][\color{red}]{@}{@},
}

\lstdefinelanguage{json}{
    basicstyle=\scriptsize	\ttfamily,
    numbers=left,
    numberstyle=\scriptsize,
    stepnumber=1,
    numbersep=8pt,
    showstringspaces=false,
    breaklines=true,
    frame=lines,
    numbers=none,
    backgroundcolor=\color{background},
    moredelim=**[is][\color{blue}]{@}{@},
    literate=
     *{0}{{{\color{numb}0}}}{1}
      {1}{{{\color{numb}1}}}{1}
      {2}{{{\color{numb}2}}}{1}
      {3}{{{\color{numb}3}}}{1}
      {4}{{{\color{numb}4}}}{1}
      {5}{{{\color{numb}5}}}{1}
      {6}{{{\color{numb}6}}}{1}
      {7}{{{\color{numb}7}}}{1}
      {8}{{{\color{numb}8}}}{1}
      {9}{{{\color{numb}9}}}{1}
      {:}{{{\color{punct}{:}}}}{1}
      {,}{{{\color{punct}{,}}}}{1}
      {\{}{{{\color{delim}{\{}}}}{1}
      {\}}{{{\color{delim}{\}}}}}{1}
      {[}{{{\color{delim}{[}}}}{1}
      {]}{{{\color{delim}{]}}}}{1},
}

\lstdefinestyle{yaml}{
    language=YAML,
    basicstyle=\small\ttfamily,
    keywordstyle=\color{blue},
    commentstyle=\color{gray},
    stringstyle=\color{orange},
    showstringspaces=false,
    breaklines=true,
    frame=single,
    backgroundcolor=\color{lightgray},
    moredelim=**[is][\color{blue}]{@}{@},
}

\begin{document}

\title{SAGE - A Tool for Optimal Deployments in Kubernetes Clusters
\thanks{This work was supported by a grant of the Romanian National Authority for Scientific Research and Innovation, CNCS/CCCDI - UEFISCDI, project number PN-III-P1-1.1-TE-2021-0676, within PNCDI III.}
}

\author{\IEEEauthorblockN{Vlad-Ioan Luca\IEEEauthorrefmark{1}
and
M\u{a}d\u{a}lina Era\c{s}cu\IEEEauthorrefmark{2}}
\IEEEauthorblockA{Faculty of Mathematics and Informatics, West University of Timisoara\\
4 blvd. V. Parvan, 300223, Timisoara, Romania \\
Email: \IEEEauthorrefmark{1}ioan.luca99@e-uvt.ro,
\IEEEauthorrefmark{2}madalina.erascu@e-uvt.ro
}}

\maketitle

\begin{abstract}
Cloud computing has brought a fundamental transformation in how organizations operate their applications, enabling them to achieve affordable high availability of services. Kubernetes has emerged as the preferred choice for container orchestration and service management across many Cloud computing platforms. The scheduler in Kubernetes plays a crucial role in determining the placement of newly deployed service containers. However, the default scheduler, while fast, often lacks optimization, leading to inefficient service placement or even deployment failures. 

This paper introduces SAGE, a tool for optimal solutions in Kubernetes clusters that can also assist the Kubernetes default scheduler and any other custom scheduler in application deployment. SAGE computes an optimal deployment plan based on the constraints of the application to be deployed and the available Cloud resources. We show the potential benefits of using SAGE by considering test cases with various characteristics. It turns out that SAGE  surpasses other schedulers by comprehensively analyzing the application demand and cluster image. This ability allows it to better understand the needs of the pods, resulting in consistently optimal solutions across all scenarios.
The accompanying material of this paper is publicly available at \url{https://github.com/SAGE-Project/SAGE-Predeployer}.
\end{abstract}


\section{Introduction}\label{sec:Introduction}
The advent of cloud computing radically changed the software development. In terms of architecture, it led to the transition from monolithic software architectures to a more loosely coupled and cloud-enabled architecture such as component-based applications\footnote{https://www.gartner.com/en/documents/3624017} and micro-services \cite{balalaie2016microservices}. In terms of life cycle, for shorter timing between idea and deployment, it facilitated the development of DevOps paradigm. At the early stages of software development, \emph{application modeling} would benefit from techniques for automatic and optimal deployment, which encompasses two key aspects:
\begin{inparaenum}[\itshape (1)\upshape] 
\item the synthesis of deployment plans that are optimal by design, and
\item the integration of such deployment plans into the application modeling process, enabling formal reasoning on a model of the deployed application.
\end{inparaenum}

The process of \emph{automated deployment} typically involves:
\begin{inparaenum}[\itshape (1)\upshape] 
\item selecting computing resources, 
\item distributing or assigning application components across available resources, and 
\item dynamically modifying the deployment to handle peaks in user requests. 
\end{inparaenum} 

In \cite{ERASCU2021100664}, we tackled the first two steps by developing a recommendation engine. In this paper, we show the benefits of using such a recommendation engine in the deployment of service containers managed by Kubernetes~\cite{Kubernetes}, abreviated as K8s. More precisely, we developed SAGE, which at the initial stage of deployment, can obtain significant cost savings for a K8s cluster while also helping with the selection of the most optimal hardware from a cloud hosting service in an automatic manner. Additionally, it obtains a deployment plan in scenarios where the K8s scheduler fails. SAGE integrates the recommendation engine from \cite{ERASCU2021100664}, here named SAGEOpt, and a new component named SAGE Predeployer. We chose K8s as it has emerged, in the last years, as the industry standard for container orchestration and service management supporting DevOps CI/CD paradigm. K8s is an open-source project that acts as a bridge between the cluster operator and the applications running on the cluster. Applications are structured as sets of services, each developed, deployed, and scaled independently.

SAGE takes as input an application description (components and their hardware/software requirements and the constraints in between) and the list of available Cloud Provides offers, and uses SAGEOpt to output, in an automatic manner, an optimal deployment plan, i.e. assignment of components on offers and the type of these offers, such that the application constraints are fulfilled and the leasing costs are minimized. SAGE Predeployer translates the SAGEOpt deployment plan into specific K8s constructs which are used for creating manifest files to be deployed to the K8s cluster.

By using various test cases, we show that SAGE can be used in scenarios with typical K8s deployment constraints: pod affinity, anti-affinity, anti-affinity to itself, number of replicas and node affinity. Moreover, SAGE excels in scenarios where an analysis of the entire deployment is needed to determine the optimal solution. In such cases, it successfully determines optimal pod placements when the default scheduler of K8s and of Boreas \cite{boreas-paper}, a scheduler improving the K8s one, fail. These findings highlight SAGE as a cost-effective and reliable approach that reduces infrastructure expenses by optimizing the pre-deployment phase and generating an optimal solution all by maintaining the application requirements.

\section{Related Work}\label{sec:RelatedWork}
SAGE can be used for the selection of the computing resources and the distribution/assignment of the application components over the available computing resources but to be used during the deployment, as a resource scheduler, it has to be further extended. At this stage, it can also be used as a complementary tool for a K8s scheduler as a deployment orchestrator.

The most similar works related to our are \cite{DBLP:conf/setta/AbrahamCJKM16,de2019modeling,lebesbye2021boreas}. Paper \cite{DBLP:conf/setta/AbrahamCJKM16} introduces Zephyrus2 which has a role similar to the SAGEOpt. Different from our approach, in Zephyrus2, the list of offers is limited (only four types of VMs) while in our case it is dynamically constructed based on the cloud providers (CPs) offers. Zephyrus2 can handle a similar set of constraints in between components as ours, and, additionally, binding preferences allowing the connection of components with components located in the same availability zone or region. Zephyrus2 characterizes CP offers only by memory and price requirements but their approach can be easily extended to other types of resources. Binding preferences can easily be implemented in SAGEOpt.  

Paper \cite{de2019modeling} used Zephyrus2 to synthesize an initial static deployment (an approach similar to SAGE), and subsequently, to synthesize dynamic deployment actions. Dynamic deployment actions represent auto-healing (in case of faults) and scaling in and out instances of the components. This was applied to one complex case study The Fredhopper Cloud Services, now part of \url{https://www.attraqt.com/}. However, there is no study on the benefit of an initial deployment as compared to the case it is missing.

Zephyrus2 is also used in Boreas \cite{lebesbye2021boreas}, a custom K8s scheduler that can be used to deploy service containers optimally in terms of price such that the constraints between containers is maintained. Different from our approach, Boreas does not use an initial optimal placement of pods and uses the entire cluster of nodes. Different from SAGE, it can be used also as a resource scheduler. K8s \cite{Kubernetes} also has a default scheduler for service container deployment but, it does not use an initial optimal placement of pods and, differently to Boreas, does not optimally place them hence it might fail when Boreas does not. Except for Boreas, there are also other approaches for optimizing the K8s scheduler~\cite{9170419,gog2016firmament,han2021tailored}, however, they take into account deployment information like the type of applications deployed, order, and duration of tasks, just to name a few, and this is our aim for future work. 
\section{Preliminaries}\label{sec:Preliminaries}
\subsection{K8s Terminology}\label{K8sTerminology}
Kubernetes \cite{k8s}, abreviated K8s, is an open-source container orchestration platform that facilitates the deployment, scaling, and management of containerized applications across a cluster of computers. A \emph{cluster} consists of a set of interconnected machines, called \emph{nodes}. These nodes can be physical servers or virtual machines. Each node runs a K8s runtime environment, which enables it to host and manage \emph{containers}.

\emph{Containers} are lightweight, isolated environments that package applications and their dependencies, providing consistency and portability across different computing environments. 

To effectively manage containerized applications, K8s uses \emph{pods}. It encapsulates one or more containers that are co-located and share the same network and storage resources. Pods are scheduled onto nodes and are the atomic unit of deployment in K8s. Multiple pods are grouped together within a higher-level abstraction called a \emph{deployment}. 
A \emph{deployment} manages the lifecycle of pods, ensuring the desired number of replicas are running and providing self-healing capabilities by replacing failed pods.
Nodes, on the other hand, provide the underlying infrastructure to host and execute the pods. 
Understanding these fundamental concepts is crucial for comprehending the interaction between the SAGE Predeployer, K8s, and the deployment process.

\emph{Pod and node affinities and anti-affinities}, in K8s, are mechanisms that influence the scheduling of pods onto nodes based on specific criteria. These features provide control over pod distribution within a cluster. \emph{Pod affinity} allows to define rules for scheduling pods based on the presence of other pods. It promotes co-location of related pods on the same node or their spread across different nodes, improving locality and reducing network latency. \emph{Pod anti-affinity} ensures that the specified pods do not run on the same node, increasing fault tolerance and high availability. 
A deployment can have \emph{anti-affinity with itself} to ensure resiliency. \emph{Node affinity} influences pod scheduling based on node properties such as labels or resource capacities, optimizing workload placement. We will use these mechanisms to assign pods to the required nodes as specified by the SAGEOpt.


\subsection{K8s Scheduler}\label{sec:K8sScheduler}

K8s uses a scheduler to find a suitable place for a pod on a nod. To do this, it follows these steps:
\begin{inparaenum}[\itshape (1)\upshape] 
\item\emph{Pod Watching}: The scheduler continuously monitors the K8s API server for newly created pods without assigned nodes.
\item\emph{Filtering (Predicates)}: The scheduler applies filter functions to some nodes, checking if each node can accommodate the pod based on available resources, node conditions, taints/tolerations, and policy constraints.
\item\emph{Scoring (Priorities)}: The scheduler assigns a score to each feasible node using scoring functions. 
\item\emph{Node Selection}: The node with the highest total score is selected to run the pod.
\item\emph{Pod Binding}: The selected node is communicated to the system and the pod is placed on it.
\end{inparaenum}

One of the main concerns of the K8s scheduler is the speed of execution and, because of this, the scheduler implements a series of optimizations that make it hard to find the optimal solution. For example, the node threshold\footnote{https://kubernetes.io/docs/concepts/scheduling-eviction/scheduler-perf-tuning/} forces the scheduler to look only at a percentage of nodes when taking the decision where to place the pod. Another one is the fact that the scheduler does not analyze the whole batch of manifest files received when placing a pod. These might have negative implications, see Section~\ref{sec:ExperimentalAnalysis}.  

\section{SAGE - A Tool for Optimal Deployments in Kubernetes Clusters}\label{sec:SAGE}
\begin{figure}
\centering

\resizebox{\columnwidth}{!}{

\tikzset{every picture/.style={line width=0.75}} 

\begin{tikzpicture}[x=0.75pt,y=0.75pt,yscale=-1,xscale=1]

\draw  [fill={rgb, 255:red, 255; green, 255; blue, 255 }  ,fill opacity=1 ] (81.4,166) -- (33.81,166) -- (33.81,80) -- (101.8,80) -- (101.8,145.6) -- cycle -- (81.4,166) ; \draw   (101.8,145.6) -- (85.48,149.68) -- (81.4,166) ;
\draw   (126.33,103.6) .. controls (126.33,75.65) and (148.99,53) .. (176.93,53) -- (471.07,53) .. controls (499.01,53) and (521.67,75.65) .. (521.67,103.6) -- (521.67,255.4) .. controls (521.67,283.35) and (499.01,306) .. (471.07,306) -- (176.93,306) .. controls (148.99,306) and (126.33,283.35) .. (126.33,255.4) -- cycle ;
\draw  [fill={rgb, 255:red, 255; green, 255; blue, 255 }  ,fill opacity=1 ] (154,155) -- (234,155) -- (234,205) -- (154,205) -- cycle ;

\draw    (101,122) -- (152.29,176.38) ;
\draw [shift={(153.67,177.83)}, rotate = 226.67] [color={rgb, 255:red, 0; green, 0; blue, 0 }  ][line width=0.75]    (10.93,-3.29) .. controls (6.95,-1.4) and (3.31,-0.3) .. (0,0) .. controls (3.31,0.3) and (6.95,1.4) .. (10.93,3.29)   ;
\draw    (89,238.67) -- (152.21,179.2) ;
\draw [shift={(153.67,177.83)}, rotate = 136.75] [color={rgb, 255:red, 0; green, 0; blue, 0 }  ][line width=0.75]    (10.93,-3.29) .. controls (6.95,-1.4) and (3.31,-0.3) .. (0,0) .. controls (3.31,0.3) and (6.95,1.4) .. (10.93,3.29)   ;
\draw   (291,125.2) .. controls (291,105.39) and (307.06,89.33) .. (326.87,89.33) -- (475.8,89.33) .. controls (495.61,89.33) and (511.67,105.39) .. (511.67,125.2) -- (511.67,232.8) .. controls (511.67,252.61) and (495.61,268.67) .. (475.8,268.67) -- (326.87,268.67) .. controls (307.06,268.67) and (291,252.61) .. (291,232.8) -- cycle ;
\draw    (233,179) -- (289,178.03) ;
\draw [shift={(291,178)}, rotate = 179.01] [color={rgb, 255:red, 0; green, 0; blue, 0 }  ][line width=0.75]    (10.93,-3.29) .. controls (6.95,-1.4) and (3.31,-0.3) .. (0,0) .. controls (3.31,0.3) and (6.95,1.4) .. (10.93,3.29)   ;
\draw  [fill={rgb, 255:red, 255; green, 255; blue, 255 }  ,fill opacity=1 ] (298,154) -- (378,154) -- (378,204) -- (298,204) -- cycle ;

\draw  [dash pattern={on 0.84pt off 2.51pt}]  (254,181) -- (254,219) ;
\draw  [fill={rgb, 255:red, 255; green, 255; blue, 255 }  ,fill opacity=1 ] (266.67,302) -- (223,302) -- (223,223) -- (285.39,223) -- (285.39,283.28) -- cycle -- (266.67,302) ; \draw   (285.39,283.28) -- (270.41,287.03) -- (266.67,302) ;
\draw  [fill={rgb, 255:red, 255; green, 255; blue, 255 }  ,fill opacity=1 ] (85.4,287) -- (37.81,287) -- (37.81,201) -- (105.8,201) -- (105.8,266.6) -- cycle -- (85.4,287) ; \draw   (105.8,266.6) -- (89.48,270.68) -- (85.4,287) ;
\draw  [fill={rgb, 255:red, 255; green, 255; blue, 255 }  ,fill opacity=1 ] (452.2,134.33) -- (413,134.33) -- (413,61.33) -- (469,61.33) -- (469,117.53) -- cycle -- (452.2,134.33) ; \draw   (469,117.53) -- (455.56,120.89) -- (452.2,134.33) ;
\draw  [fill={rgb, 255:red, 255; green, 255; blue, 255 }  ,fill opacity=1 ] (452.87,297) -- (413.67,297) -- (413.67,224) -- (469.67,224) -- (469.67,280.2) -- cycle -- (452.87,297) ; \draw   (469.67,280.2) -- (456.23,283.56) -- (452.87,297) ;
\draw  [fill={rgb, 255:red, 255; green, 255; blue, 255 }  ,fill opacity=1 ] (452.87,215.67) -- (413.67,215.67) -- (413.67,142.67) -- (469.67,142.67) -- (469.67,198.87) -- cycle -- (452.87,215.67) ; \draw   (469.67,198.87) -- (456.23,202.23) -- (452.87,215.67) ;
\draw    (379,178) -- (412.19,103.16) ;
\draw [shift={(413,101.33)}, rotate = 113.92] [color={rgb, 255:red, 0; green, 0; blue, 0 }  ][line width=0.75]    (10.93,-3.29) .. controls (6.95,-1.4) and (3.31,-0.3) .. (0,0) .. controls (3.31,0.3) and (6.95,1.4) .. (10.93,3.29)   ;
\draw    (379,178) -- (413,178.63) ;
\draw [shift={(415,178.67)}, rotate = 181.06] [color={rgb, 255:red, 0; green, 0; blue, 0 }  ][line width=0.75]    (10.93,-3.29) .. controls (6.95,-1.4) and (3.31,-0.3) .. (0,0) .. controls (3.31,0.3) and (6.95,1.4) .. (10.93,3.29)   ;
\draw    (379,178) -- (412.89,258.82) ;
\draw [shift={(413.67,260.67)}, rotate = 247.25] [color={rgb, 255:red, 0; green, 0; blue, 0 }  ][line width=0.75]    (10.93,-3.29) .. controls (6.95,-1.4) and (3.31,-0.3) .. (0,0) .. controls (3.31,0.3) and (6.95,1.4) .. (10.93,3.29)   ;
\draw   (544.8,165.14) .. controls (543.9,159.15) and (546.84,153.21) .. (552.38,149.86) .. controls (557.91,146.5) and (565.07,146.31) .. (570.81,149.37) .. controls (572.84,145.89) and (576.56,143.48) .. (580.84,142.88) .. controls (585.13,142.28) and (589.47,143.56) .. (592.56,146.33) .. controls (594.29,143.17) and (597.69,141.05) .. (601.55,140.71) .. controls (605.41,140.38) and (609.19,141.88) .. (611.54,144.69) .. controls (614.67,141.34) and (619.65,139.93) .. (624.32,141.07) .. controls (629,142.21) and (632.52,145.69) .. (633.38,150.02) .. controls (637.22,150.97) and (640.41,153.39) .. (642.14,156.64) .. controls (643.86,159.9) and (643.96,163.69) .. (642.39,167.02) .. controls (646.17,171.48) and (647.05,177.44) .. (644.71,182.66) .. controls (642.37,187.88) and (637.17,191.58) .. (631.03,192.37) .. controls (630.99,197.27) and (628.03,201.77) .. (623.31,204.12) .. controls (618.59,206.48) and (612.83,206.34) .. (608.26,203.74) .. controls (606.31,209.61) and (600.83,213.93) .. (594.18,214.83) .. controls (587.53,215.73) and (580.91,213.05) .. (577.18,207.96) .. controls (572.6,210.47) and (567.11,211.19) .. (561.94,209.97) .. controls (556.77,208.74) and (552.36,205.66) .. (549.7,201.43) .. controls (545.02,201.93) and (540.5,199.73) .. (538.38,195.91) .. controls (536.25,192.1) and (536.98,187.49) .. (540.2,184.37) .. controls (536.03,182.13) and (533.9,177.7) .. (534.92,173.38) .. controls (535.95,169.06) and (539.89,165.83) .. (544.7,165.37) ; \draw   (540.2,184.37) .. controls (542.17,185.42) and (544.44,185.9) .. (546.72,185.74)(549.7,201.43) .. controls (550.68,201.33) and (551.64,201.11) .. (552.55,200.78)(577.18,207.96) .. controls (576.49,207.02) and (575.91,206.01) .. (575.46,204.96)(608.26,203.74) .. controls (608.61,202.67) and (608.84,201.57) .. (608.94,200.46)(631.03,192.37) .. controls (631.08,187.16) and (627.82,182.38) .. (622.66,180.1)(642.39,167.02) .. controls (641.55,168.79) and (640.28,170.37) .. (638.66,171.62)(633.38,150.02) .. controls (633.52,150.73) and (633.59,151.46) .. (633.58,152.19)(611.54,144.69) .. controls (610.76,145.53) and (610.12,146.46) .. (609.63,147.46)(592.56,146.33) .. controls (592.14,147.09) and (591.83,147.89) .. (591.63,148.72)(570.81,149.37) .. controls (572.02,150.02) and (573.14,150.8) .. (574.15,151.69)(544.8,165.14) .. controls (544.92,165.97) and (545.12,166.78) .. (545.38,167.58) ;
\draw    (469.5,97) -- (532.76,177.43) ;
\draw [shift={(534,179)}, rotate = 231.81] [color={rgb, 255:red, 0; green, 0; blue, 0 }  ][line width=0.75]    (10.93,-3.29) .. controls (6.95,-1.4) and (3.31,-0.3) .. (0,0) .. controls (3.31,0.3) and (6.95,1.4) .. (10.93,3.29)   ;
\draw    (470.5,179) -- (532,179) ;
\draw [shift={(534,179)}, rotate = 180] [color={rgb, 255:red, 0; green, 0; blue, 0 }  ][line width=0.75]    (10.93,-3.29) .. controls (6.95,-1.4) and (3.31,-0.3) .. (0,0) .. controls (3.31,0.3) and (6.95,1.4) .. (10.93,3.29)   ;
\draw    (471,261) -- (532.78,180.59) ;
\draw [shift={(534,179)}, rotate = 127.53] [color={rgb, 255:red, 0; green, 0; blue, 0 }  ][line width=0.75]    (10.93,-3.29) .. controls (6.95,-1.4) and (3.31,-0.3) .. (0,0) .. controls (3.31,0.3) and (6.95,1.4) .. (10.93,3.29)   ;

\draw (145,66.67) node [anchor=north west][inner sep=0.75pt]   [align=left] {SAGE};
\draw (301,98.33) node [anchor=north west][inner sep=0.75pt]   [align=left] {{\small SAGE }\\{\small Predeployer}};
\draw (164,170.5) node [anchor=north west][inner sep=0.75pt]   [align=left] {{\small SAGEOpt}};
\draw (303,169.5) node [anchor=north west][inner sep=0.75pt]   [align=left] {{\small Translator}};
\draw (228,240.55) node [anchor=north west][inner sep=0.75pt]   [align=left] {\begin{minipage}[lt]{36.9pt}\setlength\topsep{0pt}
\begin{center}
{\small Optimal }\\{\small Solution}
\end{center}

\end{minipage}};
\draw (33.4,99.39) node [anchor=north west][inner sep=0.75pt]   [align=left] {\begin{minipage}[lt]{48.64pt}\setlength\topsep{0pt}
\begin{center}
{\small App }\\{\small Description}
\end{center}

\end{minipage}};
\draw (49.9,221.03) node [anchor=north west][inner sep=0.75pt]   [align=left] {\begin{minipage}[lt]{29.26pt}\setlength\topsep{0pt}
\begin{center}
{\small Cloud }\\{\small Offers}
\end{center}

\end{minipage}};
\draw (415.24,67.69) node [anchor=north west][inner sep=0.75pt]   [align=left] {\begin{minipage}[lt]{37.42pt}\setlength\topsep{0pt}
\begin{center}
{\small SAGE}\\{\small manifest}\\{\small files}
\end{center}

\end{minipage}};
\draw (415.9,230.36) node [anchor=north west][inner sep=0.75pt]   [align=left] {\begin{minipage}[lt]{37.42pt}\setlength\topsep{0pt}
\begin{center}
{\small Boreas}\\{\small manifest}\\{\small files}
\end{center}

\end{minipage}};
\draw (415.9,149.03) node [anchor=north west][inner sep=0.75pt]   [align=left] {\begin{minipage}[lt]{37.42pt}\setlength\topsep{0pt}
\begin{center}
{\small K8s}\\{\small manifest}\\{\small files}
\end{center}

\end{minipage}};
\draw (554,168.5) node [anchor=north west][inner sep=0.75pt]   [align=left] {{\small K8s Cluster}};

\end{tikzpicture}}
\caption{SAGE General Architecture}
\label{fig:sage_arhitecture}
\end{figure}
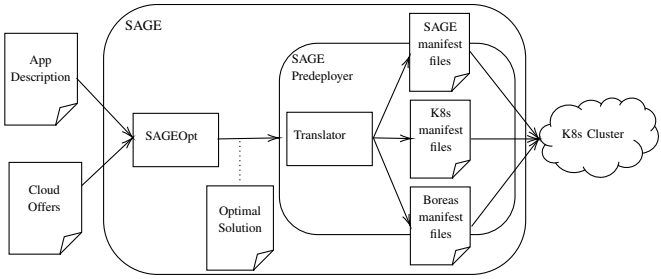

In this section we present SAGE tool (see Figure~\ref{fig:sage_arhitecture}), pipelining an optimization engine (SAGEOpt) and a predeployer (SAGE Predeployer).
\subsection{SAGE Optimization Engine}\label{sec:SAGE-Opt}
\emph{SAGE Optimization Engine (SAGEOpt)} was developed in~\cite{ERASCU2021100664}. It solves the following problem: given a description of an application (components hardware requirements and the constraints in between) and a list of offers from Cloud Providers, find, in an automatic manner, an optimal deployment plan, that is, the allocation of components on virtual machines (VMs) such that the cost is minimal and the constraints are satisfied. SAGEOpt supports a large set of constraints between components:
\begin{inparaenum}[\itshape (1)\upshape] 
    \item \noindent\emph{Conflict.} The set of conflictual components can not be deployed on the same VM. 
    \item \noindent\emph{Co-location.} The set of collocated components should not be deployed on the same VM. 
    \item \noindent\emph{Exclusive deployment.} From the set of exclusive deployment components, only one should be deployed. 
    \item \noindent\emph{Require-Provide.} Such a constraint determines the number of instances corresponding to the interacting components as follows: 
    \begin{inparaenum}[\itshape (i)\upshape]
	\item $C_i$ requires (consumes) at least $n_{ij}$ instances of $C_j$ and 
	\item $C_j$ can serve (provides) at most $m_{ij}$ instances of $C_i$.
    \end{inparaenum}
    \item \noindent\emph{Full deployment.} The component being in full deployment must be deployed on all leased VMs (except on those which would induce conflicts on components).
    \item \noindent\emph{Deployment with bounded number of instances.} The number of instances corresponding to deployed component should be equal, greater or less than some values. Implicitly, the instances of the same component are assigned to different VMs in order to ensure resiliency.
\end{inparaenum}

\subsection{SAGE Predeployer}\label{sec:SAGEPredeployer}

The SAGE Predeployer is the bridge between the output of SAGEOpt and the K8s infrastructure. Its primary function is to translate the optimal solution obtained from SAGEOpt into K8s manifest files and facilitate their deployment to a K8s cluster (Digital Ocean\footnote{\url{https://www.digitalocean.com}} for now). To this end, it proceeds as follows:
\begin{inparaenum}[\itshape (1)\upshape]
\item Receive input
\item Translate solution
\item Create manifest files
\item Deploy to Cloud infrastructure.
\end{inparaenum}


\noindent \emph{\textbf{Step 1: Receive Input}} SAGE Predeployer calls SAGEOpt and takes as input the output produced by it. This file has two sections (see Listing ~\ref{json:input_predeployer}). The first includes the description of the application deployed (name, components and their requirements and the restrictions in between). The second, introduced by \texttt{output}, includes the optimal solution obtained by SAGEOpt: the minimal price of the solution, the type of VMs to be acquired and their specification, the allocation of components on VMs.
\begin{lstlisting}[language=json, caption={SAGE Predeployer Input file }, label={json:input_predeployer}]
{
    "application": "SecureWebContainer",
    "components": [
        {
            "id": 1,
            "name": "Balancer",
            "Compute": {
                "CPU": 1000,
                "Memory": 2048
            }, ... },
            "operatingSystem": "httpd"
        }, ...
    ],
    "restrictions": [
        {
            "type": "Conflicts",
            "alphaCompId": 1,
            "compsIdList": [2, 3, 4, 5]
        }, ...
    ],
    "output": {
        "min_price": 3360,
        "types_of_VMs": [7, 24, 10, 10, 10],
        "VMs_specs": [
            {
                "s-2vcpu-4gb": {
                    "cpu": 1800,
                    "memory": 3150,
                    "storage": 69000,
                    "operatingSystem": "nginx",
                    "price": 240,
                    "id": 7
                }
            },...
        ],
        "assign_matr": [[1, 0, 0, 0, 0],
                        [0, 0, 1, 0, 0],
                        [0, 0, 0, 1, 1],
                        [0, 1, 0, 0, 0],
                        [0, 0, 1, 1, 1]]
    }
}
\end{lstlisting}
\noindent \emph{\textbf{Step 2: Translate Solution}} SAGE Predeployer translates the optimal solution provided by the SAGEOpt into K8s manifest files. During this process, each component listed in the input file as a component is converted into a K8s deployment manifest. The relevant fields from the input file are translated as follows:
\begin{inparaenum}[\itshape (1)\upshape] 
\item \texttt{application}: will be the name of the deployment.
\item \texttt{compute}: is the hardware and software requirements of each pod.
\item \texttt{restrictions}: all restrictions of type conflict, co-location, and full deployment will be translated into pod anti-affinity, pod affinity, and pod anti-affinity to itself, respectively. 
The last constraint places each component instance (replica) on different nodes (resiliency). 
\item \texttt{types\_of\_vms} and \texttt{vms\_specs} specify the type of the nodes used and their hardware/software specifications to be used in the scheduling. In fact, these are the nodes that give the optimal solution as computed by SAGEOpt. 
\item \texttt{assign\_matr}: is a matrix with entries in $\{0,1\}$ that determines how instances of components are placed on VMs or nodes. Each column represents a VM from \texttt{types\_of\_VMs} and \texttt{VMs\_specs}, and each row represents a component. 
\end{inparaenum} 


\noindent \emph{\textbf{Step 3: Create Manifest Files}}: The SAGE Predeployer generates K8s manifest files based on the translated solution. The manifest files used by SAGE use all the information given by SAGEOpt used for deployment (see Listing~\ref{yml:manifest_SAGE}). This information includes pod affinity and anti-affinity, pod anti-affinity to itself, pod number of replicas, node affinity (information found in \texttt{assign\_matr}) and node characteristics (information found in \texttt{VMs\_specs}). 

\begin{lstlisting}[language=json, caption={Manifest file generated for deployment using SAGE for pod Balancer}, label={yml:manifest_SAGE}]
apiVersion: apps/v1
kind: Deployment
metadata:
  labels:
    app: balancer
    id: '1'
  name: balancer
spec:
  replicas: 1
  ...
  template:
    ...
    spec:
      affinity:
        @nodeAffinity:@
          @...@
            @nodeSelectorTerms:@
            @- matchExpressions:@
            @  - key: index@
            @    operator: In@
            @    values:@
            @    - '0'@
        podAntiAffinity:
          ...
          - labelSelector:
              matchExpressions:
              - key: app
                operator: In
                values:
                - apache
            topologyKey: kubernetes.io/hostname
          # all the other podAntiAffinity, in this case with nginx, idsserver, idsagent
      containers:
      - image: k8s.gcr.io/pause:2.0
        name: balancer-container
        resources:
          requests:
            cpu: 1000m
            ephemeral-storage: 0Gi
            memory: 2048Mi
\end{lstlisting}
Manifest files used by K8s (see Listing~\ref{yml:manifest_K8s}) are obtained from SAGE ones except the node affinity information as highlighted in Listing~\ref{yml:manifest_SAGE}. Compared to manifest files for K8s, Boreas ones need to substract the resources allocated to the Boreas scheduler ($100m CPU / number\_of\_all\_instances$) from the hardware requirements of each pod and also specify that the custom scheduler is used. Neither node affinities are given to the manifest files for K8s and Boreas because we want to evaluate the capabilities of their schedulers, nor pod anti-affinities with itself unless they are specified precisely in the application description file. The information included is:
\begin{inparaenum}[\itshape (1)\upshape] 
\item node types to be used in the deployment and  
\item pod affinity/anti-affinity and number of replicas.
\end{inparaenum}
\begin{lstlisting}[language=json, caption={Manifest file generated for deployment using K8s for pod Balancer}, label={yml:manifest_K8s}]
apiVersion: apps/v1
kind: Deployment
metadata:
  labels:
    app: balancer
    id: '1'
  name: balancer
spec:
  replicas: 1
  ...
  template:
    ...
    spec:
      affinity:
        podAntiAffinity:
          ...
          - labelSelector:
              matchExpressions:
              - key: app
                operator: In
                values:
                - apache
            topologyKey: kubernetes.io/hostname
          # all the other podAntiAffinity, in this case with nginx, idsserver, idsagent
      containers:
      - image: k8s.gcr.io/pause:2.0
        name: balancer-container
        resources:
          requests:
            cpu: 1000m
            ephemeral-storage: 0Gi
            memory: 2048Mi
\end{lstlisting}

\begin{lstlisting}[language=json, caption={Manifest file generated for deployment using Boreas for pod Balancer}, label={yml:manifest_Boreas}]
...
        name: balancer-container
        resources:
          requests:
            @cpu: 980m@
            ephemeral-storage: 0Gi
            memory: 2048Mi
      @schedulerName: boreas-scheduler@
\end{lstlisting}
\noindent \emph{\textbf{Step 4: Deploy to Digital Ocean}}: Once the manifest files are created, SAGE Predeployer deploys the translated SAGE solution to the Digital Ocean infrastructure. It interacts with the cloud provider's CLI to create the required resources, such as nodes and pods, according to the specified configurations. 


\section{Test Cases}\label{sec:CaseStudies}
In this section, we present test cases that will be used in Section \ref{sec:ExperimentalAnalysis} to show the potential benefits of SAGE 
\begin{inparaenum}[\itshape (i)\upshape]
\item of being used for the initial deployment plan as well as
\item compared to Boreas and K8s schedulers in offering optimal deployment plans.
\end{inparaenum} These test cases are either taken from~\cite{ERASCU2021100664} (see Sections \ref{sec:SecureBillingEmailService}--\ref{sec:oryx2}), or are proposed by us based on the conclusions drawn from experimental analysis (see Sections~\ref{sec:BatchAnalysisTest}--\ref{sec:NodeAnalysisTest}) or are those proposed by developers of Boreas (see Section~\ref{sec:BoreasPaperTests}) More tests and their analysis are available \href{https://docs.google.com/spreadsheets/d/1kLXDmJzOg0XM5SmFr2R-Tt_RecWdWB_ouR-q4RTnUQo/edit?usp=sharing}{here}.
\subsection{Secure Billing Email Service}\label{sec:SecureBillingEmailService}
In the context of a web application ensuring a secure billing email service 
we consider an architecture consisting of $5$ components: 
\begin{inparaenum}[\itshape (i)\upshape]
	\item a coding service ($C_1$), 
	\item a software manager of the user rights and privileges ($C_2$), 
	\item a gateway component ($C_3$), 
	\item an SQL server ($C_4$) and 
	\item a load balancer ($C_5$).
\end{inparaenum} 
Component $C_1$ should use exclusively a virtual machine, thus it can be considered in \emph{conflict} with all the other components. In such a case the original optimization problem can be decomposed in two subproblems, one corresponding to component $C_1$ and the other one corresponding to the other $4$ components. The first problem is trivial: find the VM with the smallest price which satisfies the hardware requirements of component $C_1$. 

The load balancing component should not be placed on the same machine as the gateway component and the SQL server (\emph{Conflict} constraint). On the other hand, only one instance of components $C_1$ and $C_5$ should be deployed while the other three components could have a larger number of instances placed on different virtual machines (\emph{Deployment with a bounded number of instances} constraint, in particular, \emph{equal bound}).

\subsection{Secure Web Container}\label{sec:SecWebContainer} 
The \emph{Secure Web Container} \cite{DBLP:journals/tsc/CasolaBEMR17} 
is a service that provides:
\begin{inparaenum}[\itshape (i)\upshape]
	\item \emph{resilience} to attacks and failures, by introducing redundancy and diversity techniques, and
	\item protection from unauthorized and potentially dangerous accesses, by integrating proper \emph{intrusion detection} tools.
\end{inparaenum}
Resilience can be implemented by a set of different Web Container components and a Balancer component, which is responsible for dispatching web requests to the active web containers to ensure load balancing. In the simplest scenario, there are two Web Containers (e.g. Apache Tomcat\footnote{\url{http://tomcat.apache.org}} and Nginx). Intrusion detection is ensured by the generation of intrusion detection reports with a certain frequency. It was implemented by deploying an IDSAgent, to be installed on the resources to be protected, and an IDSServer, which collects data gathered by the IDSAgents and performs the detection activities.

\noindent The constraints between application components are as follows. For Web resilience:
	\begin{inparaenum}[\itshape (i)\upshape]
		\item Any two of the Balancer, Apache, and Nginx components cannot be deployed on the same machine (\emph{Conflict} constraint);
		\item Exactly one Balancer component has to be instantiated (\emph{Deployment with bounded number of instances} constraint, in particular \emph{equal bound}).
		\item The total number of instances for Apache and Nginx components must be at least $3$ (level of redundancy) (\emph{Deployment with a bounded number of instances} constraint, in particular, \emph{lower bound}).
	\end{inparaenum}
For Web intrusion detection:
	\begin{inparaenum}[\itshape (i)\upshape]
		\item the IDSServer component needs exclusive use of machines (\emph{Conflict} constraint).
		\item There must be an IDSServer component additional instance every 10 IDSAgent component instances (\emph{Require-Provide} constraint).
		\item One instance of IDSAgent must be allocated on every acquired machine except where an IDSServer or a Balancer is deployed (\emph{Full Deployment} constraint).
	\end{inparaenum}
\subsection{Oryx2 application}\label{sec:oryx2}
The main goal of Oryx2 is to take incoming data and use them to create and instantiate predictive models for various use cases, e.g. movie recommendations.
It is comprised of several technologies. Both the batch and serving layer are based on Apache Spark which in turn uses both Apache Yarn\footnote{\url{http://hadoop.apache.org/}} for scheduling and Apache HDFS as a distributed file system. For a processing pipeline, Oryx2 uses Apache Kafka with at least two topics; one for incoming data and one for model update. Apache Zookeeper\footnote{\url{https://zookeeper.apache.org/}} is used by Kafka for broker coordination. All of the aforementioned technologies have subservices with a minimum system requirement and recommended deployment.
The constraints corresponding to the interactions between the components are described in the following.
\begin{inparaenum}[\itshape (i)\upshape]
\item Components HDFS.DataNode and Spark.Worker must the deployed on the same VM  (\emph{Co-location}). In this scenario, we also collocated Yarn.NodeManager because we used Yarn as a scheduler for Spark jobs.  
\item Components Kafka and Zookeeper, HDFS.NameNode and HDFS.SecondaryNameNode, YARN.ResourceManagement and HDFS.NameNode, HDFS.SecondaryNameNode, YARN.Histo\-ryService are, respectively, in \emph{conflict}, that is, they must not be placed on the same VM.
\item Components HDFS.DataNode, YARN.NodeManager and Spark.Worker must be deployed on all VMs except those hosting conflicting components  (\emph{Full Deployment}).
\item In our deployment, we consider that for one instance of Kafka there must be deployed exactly 2 instances of Zookeeper  (\emph{Require-Provide} constraint). There can be situations, however, when more Zookeeper instances are deployed for higher resilience.
\item A single instance of YARN.HistoryService, respectively Spark.HistoryService should be deployed (\emph{Deployment with bounded number of instances}  constraint, in particular \emph{equal bound}).
\end{inparaenum}
\subsection{Boreas Paper Tests}\label{sec:BoreasPaperTests}
The Boreas Paper Tests is a collection of the four tests from~\cite[Chapter~4]{boreas-master-thesis} and were focused on evaluating the impact of the Boreas scheduler on Kubernetes' deployment performance, specifically examining the differences in pod placement compared to the default scheduler. In this paper, we want to check the behavior of SAGE on these benchmarks. For the lack of space, we consider only \emph{Test D: free node with deployment constraints} (see Table~\ref{tab:Spec_configuration_test_D}).
\begin{table}[h] \centering
\begin{tabular}{|c|c|c|c|c|}
\hline
\textbf{Pod}       & \textbf{Replicas}  & \textbf{Constraints} \\ \hline
Asperitas & 3     & Anti-affinity to itself\\ \hline
Cirrus    & 2     &         \\ \hline
Cumulus   & 3     & Anti-affinity to itself, affinity to Asperitas \\ \hline
Nimbus    & 2     & Anti-affinity to itself and Asperitas\\ \hline
Stratus   & 4     &  Anti-affinity to itself\\ \hline
\end{tabular}
\caption{Specification of Test D: free node with deployment constraints}
\label{tab:Spec_configuration_test_D}
\end{table}
\subsection{Batch Analysis Test}\label{sec:BatchAnalysisTest}
The goal of this scenario, motivated by the analysis of Secure Web Container (see Section~\ref{sec:AnalysisSecureWebContainer}), is to assess the ability of each scheduler to analyze requirements of other components in the same batch (same K8s API request), in particular, it evaluates if the scheduler can anticipate the resource needs of upcoming components that are simultaneously submitted to the K8s API. We have three components: two components with a CPU request of 500m each, and one component with a CPU request of 1000m. The objective of this test is to reveal that, if the first two pods are not placed on the same node, the third pod cannot be deployed. To accomplish this, we ensure that the scenario runs on two nodes, both with 2vCPU. Neither memory considerations nor constraints between components are relevant in this specific scenario.
\subsection{Node Analysis Test}\label{sec:NodeAnalysisTest}
This test consists of three components and assesses the scheduler's ability to choose the optimal node for a pod, considering the requirements of other pods in the same K8s API batch. The inspiration behind this test is also the result from Section~ \ref{sec:AnalysisSecureWebContainer} where the scheduler was not able to better match the pod's requirements with nodes specification. This scenario involves two pods of 500m CPU and one of 2900m CPU. No constraints are present between the pods. By allocating two nodes (of 2vCPU and 4vCPU), one capable of supporting the third component and another that cannot, we evaluate the scheduler's effectiveness in node selection based on resource availability since a placement mistake of any of the first two nodes will result in a failure for the third one.

\section{Experimental Analysis}\label{sec:ExperimentalAnalysis}
The main objective of this study is to analyze how pods are placed during the initial deployment phase using three different approaches: SAGE, K8s, and Boreas. To perform the analysis, we followed a specific \emph{methodology}. We used SageOpt to obtain the optimal deployment of the application. The precise solution provided by SAGEOpt is used to create the manifest files for deployment using SAGE: pod affinity, anti-affinity and anti-affinity with itself, as well as node affinity. Notably, when dealing with K8s and Boreas manifest files, we utilized the same information for both, with one exception: the node affinity information, because the task at hand is to observe their ability to locate suitable nodes. 

To create a hardware context for the study, we deployed nodes that were identified as the most optimal by SageOpt. This was done to evaluate the performance of the schedulers in finding the best solution. We utilized the Digital Ocean offers to provide a pool of available resources from which SageOpt could choose. It is worth mentioning that the Kubernetes cluster default processes use a part of the resources available on the pod. 

The case studies considered allow us to examine how the schedulers perform during the initial deployment phase by analyzing pod placements based on the chosen hardware context.

In terms of \emph{metrics}, our focus was primarily on the placement of pods on nodes. Specifically, we evaluated whether each scheduler is capable of identifying the most optimal solution based on the recommended hardware provided by SAGEOpt. Additionally, we will assess the resilience of the solutions generated by each scheduler.

It is important to note that we will not consider scheduling time as a metric in this study. Our main objective is to identify a scheduler that minimizes deployment costs rather than prioritizing scheduling speed. Therefore, our evaluation will primarily revolve around the effectiveness and cost efficiency of the scheduler's decision-making processes, rather than the time taken for scheduling.
\subsection{Analysis: Secure Billing Email Service}\label{sec:AnalysisSecureBillingEmailService}

In this section, we look at the node placements obtained after running all three solutions in the context described in Section~\ref{sec:SecureBillingEmailService}. In Tables~\ref{tab:k8s_secure_billing} and ~\ref{tab:sage_secure_billing} 
we can observe that they are able to find a suitable solution, in particular SAGE and Boreas came up with the same pods placement. This scenario can be used as a validation that all three are set up correctly and are working properly. 

\begin{table}[htb] \centering
\begin{tabular}{|c|c|c|c|}
\hline
\backslashbox{\textbf{Pod}}{\textbf{Node}} & 8vcpu-16gb & s-8vcpu-16gb & s-8vcpu-16gb \\ \hline
Coding Service & 1 &  &  \\ \hline
Security Manager &  & 1 &  \\ \hline
Gateway &  &  & 1 \\ \hline
SQLServer &  &  & 1 \\ \hline
LoadBalancer &  & 1 &  \\ \hline
\end{tabular}
\caption{K8s solution for Secure Billing Email Service}
\label{tab:k8s_secure_billing}
\end{table}

\begin{table}[htb] \centering
\begin{tabular}{|c|c|c|c|}
\hline
\backslashbox{\textbf{Pod}}{\textbf{Node}} & 8vcpu-16gb & s-8vcpu-16gb & s-8vcpu-16gb \\ \hline
Coding Service & 1 &  &  \\ \hline
Security Manager &  &  & 1 \\ \hline
Gateway &  & 1 &  \\ \hline
SQLServer &  & 1 &  \\ \hline
LoadBalancer &  &  & 1 \\ \hline
\end{tabular}
\caption{SAGE and Boreas solutions for Secure Billing Email Service}
\label{tab:sage_secure_billing}
\end{table}

\subsection{Analysis: Secure Web Container}\label{sec:AnalysisSecureWebContainer}

This section considers the pod's placements (see Tables \ref{tab:k8s_secure_web} and  \ref{tab:sage_secure_web}) in the context of Secure Web Container from Section~\ref{sec:SecWebContainer}. One could observe that SAGE and Boreas are able to obtain (the same) optimal solution, while K8s fails. Specifically, the issue arises with the deployment of the IDSServer pod as K8s fails to properly assign a feasible node for it, likely due to the Balancer replica not being placed correctly. Since it requires only 1000m CPU, it could be accommodated on a 2vCPU node, however, it is, instead, placed on a 4vCPU node, leaving the IDSServer without a suitable node for deployment. This behavior can be explained based on the features of K8s scheduler as described in Section~\ref{sec:K8sScheduler}. More precisely, this issue could arise because K8s only analyzes the current pod it needs to deploy and does not consider the future requirements of other pods. Additionally, this can be caused due to the fact that it analyzes only up to 50\% of the total number of nodes if a suitable solution is found. This limited scope of analysis could lead to suboptimal pod placement decisions and potentially result in an increase in resource costs.
\begin{table}[h!] \centering
\resizebox{\columnwidth}{!}{\begin{tabular}{|c|c|c|c|c|c|}
\hline
\backslashbox[15mm]{\textbf{Pod}}{\textbf{Node}} & 2vcpu-4gb & 4vcpu-32gb & 4vcpu-8gb & 4vcpu-8gb & 4vcpu-8gb \\ \hline
Balancer &  & 1 &  &  &  \\ \hline
Apache &  &  & 1 &  & 1 \\ \hline
Ngnix &  &  &  & 1 &  \\ \hline
IDSServer &  &  &  &  &  \\ \hline
IDSAgent &  &  & 1 & 1 & 1 \\ \hline
\end{tabular}}
\caption{K8s solution for Secure Web Container}
\label{tab:k8s_secure_web}
\end{table}
\begin{table}[h!] \centering
\resizebox{\columnwidth}{!}{\begin{tabular}{|c|c|c|c|c|c|}
\hline
\backslashbox[15mm]{\textbf{Pod}}{\textbf{Node}} & 2vcpu-4gb & 4vcpu-32gb & 4vcpu-8gb & 4vcpu-8gb & 4vcpu-8gb \\ \hline
Balancer & 1 &  &  &  &  \\ \hline
Apache &  &  & 1 &  & 1 \\ \hline
Ngnix &  &  &  & 1 &  \\ \hline
IDSServer &  & 1 &  &  &  \\ \hline
IDSAgent &  &  & 1 & 1 & 1 \\ \hline
\end{tabular}}
\caption{SAGE and Boreas solutions for Secure Web Container}
\label{tab:sage_secure_web}
\end{table}
\subsection{Analysis: Oryx2}\label{sec:AnalysisOryx2}

This section examines the specific scenario described in Section~\ref{sec:oryx2}. From Table~\ref{tab:boreas_oryx2}, one could observe that Boreas scheduler encounters difficulties when deploying the third replica of the Yarn.NodeManager pod (denoted by \text{\sffamily X} on column s-8vcpu-16gb). This issue arises because both instances of the Zookeeper component are placed on the same node, leaving insufficient resources for the Yarn.NodeManager pod. This is because Boreas scheduler uses as few as possible nodes when all other placement requirements are fulfilled \cite{boreas-master-thesis}. In contrast, K8s (see Table~\ref{tab:k8s_oryx2}) and SAGE (see Table~\ref{tab:sage_oryx2}) are able to find solutions in this case due to their default constraint of resiliency, which involves placing replicas of the same pod on different nodes whenever possible. Boreas scheduler could address this problem by introducing pod anti-affinity to itself for Zookeeper pod, preventing multiple instances from being placed on the same node. Since it is not specified in the Boreas paper \cite{boreas-paper} when one should or should not use the pod anti-affinity to itself, this paper does not use it by default for benchmarking the Boreas scheduler.
\begin{table}[h] \centering
\begin{tabular}{|c|c|c|c|}
\hline
\backslashbox[20mm]{\textbf{Pod}}{\textbf{Node}} & s-8vcpu-16gb& s-8vcpu-16gb& s-8vcpu-16gb\\ \hline
Kafka &  &  & 1 \\ \hline
Zookeeper & 2 &  &  \\ \hline
HDFS.NameN &  &  & 1 \\ \hline
HDFS.2NameN & 1 &  &  \\ \hline
HDFS.DataN & 1 & 1 & 1 \\ \hline
YARN.ResMng &  & 1 &  \\ \hline
Yarn.HistServ & 1 &  & 1 \\ \hline
Yarn.NodeMng & 1 & 1 & \text{\sffamily X} \\ \hline
Spark.Worker & 1 & 1 & 1 \\ \hline
Spark.HistServ &  &  & 1 \\ \hline
\end{tabular}
\caption{Boreas solution for Oryx2}
\label{tab:boreas_oryx2}
\end{table}
\begin{table}[h] \centering
\begin{tabular}{|c|c|c|c|}
\hline
\backslashbox[20mm]{\textbf{Pod}}{\textbf{Node}} & s-8vcpu-16gb& s-8vcpu-16gb & s-8vcpu-16gb\\ \hline
Kafka &  & 1 &  \\ \hline
Zookkeper & 1 &  & 1 \\ \hline
HDFS.NameN &  &  & 1 \\ \hline
HDFS.2NameN &  & 1 &  \\ \hline
HDFS.DataN & 1 & 1 & 1 \\ \hline
YARN.ResMng & 1 &  &  \\ \hline
Yarn.HistServ &  &  & 1 \\ \hline
Yarn.NodeMng & 1 & 1 & 1 \\ \hline
Spark.Worker & 1 & 1 & 1 \\ \hline
Spark.HistServ & 1 &  &  \\ \hline
\end{tabular}
\caption{K8s solution for Oryx2}
\label{tab:k8s_oryx2}
\end{table}
\begin{table}[h] \centering
\begin{tabular}{|c|c|c|c|}
\hline
\backslashbox[20mm]{\textbf{Pod}}{\textbf{Node}} & s-8vcpu-16gb & s-8vcpu-16gb & s-8vcpu-16gb \\ \hline
Kafka &  &  & 1 \\ \hline
Zookeeper & 1 &  & 1 \\ \hline
HDFS.NameN &  &  & 1 \\ \hline
HDFS.2NameN & 1 &  &  \\ \hline
HDFS.DataN & 1 & 1 & 1 \\ \hline
YARN.ResMng &  & 1 &  \\ \hline
Yarn.HistServ &  &  & 1 \\ \hline
Yarn.NodeMng & 1 & 1 & 1 \\ \hline
Spark.Worker & 1 & 1 & 1 \\ \hline
Spark.HistServ &  & 1 &  \\ \hline
\end{tabular}
\caption{SAGE solution for Oryx2}
\label{tab:sage_oryx2}
\end{table}
\subsection{Analysis: Boreas tests}\label{sec:}
For the test in Section~\ref{sec:BoreasPaperTests}, all three schedulers find a deployment (see Tables~\ref{tab:K8s_boreas_test_D}-\ref{tab:sage_boreas_test_D}). Note that the test was not performed on the hardware specified in \cite{boreas-master-thesis}, which uses their entire cluster of six nodes with the same configuration, but on the hardware that SAGEOpt deems optimal for the application from Digital Ocean infrastructure that we use in our experiments. 
\begin{table}[h!] \centering
\begin{tabular}{|c|c|c|c|c|c|}
\hline
\backslashbox[15mm]{\textbf{Pod}}{\textbf{Node}} & 2vcpu-4gb & 2vcpu-4gb & 2vcpu-4gb & 4vcpu-8gb & 4vcpu-8gb\\ \hline
asperitas & 1 & & & 1 & 1 \\ \hline
cirrus    &   & & & 1 & 1 \\ \hline
cumulus   & 1 & & & 1 & 1 \\ \hline
nimbus    &   & 1 & 1 &  &  \\ \hline
stratus   &  & 1 & 1 & 1 & 1 \\ \hline
\end{tabular}
\caption{K8s solution for Boreas Test D}
\label{tab:K8s_boreas_test_D}
\end{table}
\begin{table}[h!] \centering
\begin{tabular}{|c|c|c|c|c|c|}
\hline
\backslashbox[15mm]{\textbf{Pod}}{\textbf{Node}} & 2vcpu-4gb & 2vcpu-4gb & 2vcpu-4gb & 4vcpu-8gb & 4vcpu-8gb\\ \hline
asperitas &  & & 1 & 1 & 1 \\ \hline
cirrus    & 1  & & & 1 &  \\ \hline
cumulus   &  & & 1 & 1 & 1 \\ \hline
nimbus    & 1  & 1 &  &  &  \\ \hline
stratus   & 1 & 1 &  & 1 & 1 \\ \hline
\end{tabular}
\caption{SAGE and Boreas solution for SAGE Test D}
\label{tab:sage_boreas_test_D}
\end{table}
%
\subsection{Analysis: Batch Test}\label{sec:BatchAnalysis}
This section presents the results of running the three schedulers in the context described in Section~\ref{sec:BatchAnalysisTest}. The objective of this test is to evaluate the ability of the schedulers to analyze all components within the same batch. Specifically, we assess if the schedulers can identify that the third component requires a dedicated node while the first two components should be placed on the same node. The findings, shown in Table~ \ref{tab:batch_test}, reveal that the K8s scheduler fails to deploy the third component as it mistakenly assigns the first two components to different nodes, resulting in resource starvation for the third component. This is likely due to the K8s scheduler's limited batch analysis, which prioritizes node scoring based on available resources during the node scoring phase (see Section \ref{sec:K8sScheduler}). Boreas also fails to deploy the third pod (see Table~\ref{tab:batch_test} without a specific explanation identified especially in the light that according to the documentation, Boreas seeks to minimize the number of nodes if all constraints are satisfied, hence the first two pods should have been placed on the same node. SAGE managed to deploy all pods (see Table~\ref{tab:batch_test}).

\begin{table}[h!]
\begin{tabular}{|c|c|c|}
\hline
\backslashbox[15mm]{\textbf{Pod}}{\textbf{Node}} & 2vcpu & 2vcpu \\ \hline
P1 & 1 &  \\ \hline
P2 &  & 1 \\ \hline
P3 &  &  \\ \hline
\end{tabular}
\hfill
\begin{tabular}{|c|c|c|}
\hline
\backslashbox[15mm]{\textbf{Pod}}{\textbf{Node}} & 2vcpu & 2vcpu \\ \hline
P1 &  & 1 \\ \hline
P2 &  & 1 \\ \hline
P3 & 1 &  \\ \hline
\end{tabular}
\caption{K8s, Boreas (left) and SAGE (right) solution for Batch Test}
\label{tab:batch_test}

\end{table}


\subsection{Analysis: Node Test}\label{sec:NodeAnalysis}

This section focuses on the results obtained from conducting the test described in Section \ref{sec:NodeAnalysisTest}. The objective of the test is to determine if the scheduler can effectively match pods with the most suitable node type. Specifically, we aim to verify if the first two pods can be placed on the 2vcpu node and the third pod on the 4vcpu node, as any deviation would result in resource starvation for the third pod. Reviewing Tables \ref{tab:k8s_boreas_node_test}-, we observe that SAGE successfully finds the optimal solution. However, both Boreas and K8s fail to do so, placing the first and second pods on the 4vcpu node. The K8s failure can be attributed to its default behavior of analyzing a maximum of 50\% of the available nodes for performance optimization purposes \footnote{https://kubernetes.io/docs/concepts/scheduling-eviction/scheduler-perf-tuning/}. Regarding Boreas, the precise cause of the failure remains unclear, but similar to the previous example, it appears to choose the node with the most available resources for pod placement.
\begin{table}[h!]
\centering
\begin{tabular}{|c|c|c|}
\hline
\backslashbox[15mm]{\textbf{Pod}}{\textbf{Node}} & 2vcpu-2gb & 4vcpu-8gb \\ \hline
P1 &  & 1 \\ \hline
P2 &  & 1 \\ \hline
P3 &  &  \\ \hline
\end{tabular}
\begin{tabular}{|c|c|c|}
\hline
\backslashbox[15mm]{\textbf{Pod}}{\textbf{Node}} & 2vcpu-2gb & 4vcpu-8gb \\ \hline
P1 & 1 &  \\ \hline
P2 &  & 1 \\ \hline
P3 &  &  \\ \hline
\end{tabular}
\caption{K8s (left) and Boreas (right) solution for Node Test}
\label{tab:k8s_boreas_node_test}
\end{table}

\begin{table}[h!]
\centering
\begin{tabular}{|c|c|c|}
\hline
\backslashbox[15mm]{\textbf{Pod}}{\textbf{Node}} & 2vcpu-2gb & 4vcpu-8gb \\ \hline
P1 & 1 &  \\ \hline
P2 & 1 &  \\ \hline
P3 &  & 1 \\ \hline
\end{tabular}
\caption{SAGE solution for Node Test}
\label{tab:sage_node_test}
\end{table}

\section{Conclusion and Future Work}\label{sec:ConclFutureWork}

This study has established reliable benchmarks for infrastructure solutions and demonstrated the effectiveness of the SAGE. Through its comprehensive analysis, it consistently identifies optimal solutions, leading to improved scheduling efficiency and resource allocation.

To further advance this field we will develop a custom Kubernetes scheduler based on the SAGE approach which would handle the dynamic modification of the deployment to support peaks in user requests. Moreover, we will tackle more real-life cases. These efforts will provide practical validation of the SAGE scheduler's performance and contribute to ongoing enhancements of scheduling algorithms for better cluster management.

\bibliographystyle{abbrv}
\bibliography{mybib}

\end{document}